\documentclass[prl,twocolumn,10pt,showpacs,showkeys,preprintnumbers,amsmath,amssymb,unsortedaddress,aps]{revtex4-1}
\usepackage{graphicx}
\usepackage{amssymb}
\usepackage{amsmath}
\usepackage{textcomp}

\newcommand{\ud}{\,{\mathrm{d}}}
\newcommand{\zc}{\overline{z}}

\newcommand{\fc}{\overline{f}}

\newcommand{\uMs}{M_{\mathrm{S}}}
\newcommand{\umx}{m_{\mathrm{X}}}
\newcommand{\umy}{m_{\mathrm{Y}}}
\newcommand{\umz}{m_{\mathrm{Z}}}

\DeclareMathOperator{\Arg}{Arg}
\def\be{\begin{equation}}
\def\ee{\end{equation}}

\begin{document}

\title{Topological constraints on positions of magnetic solitons in multiply-connected planar magnetic nano-elements.}

\author{Andrei B. Bogatyr\"ev}
\affiliation{Institute for Numerical Mathematics, Russian Academy of Sciences, 8 Gubkina str., Moscow GSP-1, Russia 119991}
\email{gourmet@inm.ras.ru}

\author{Konstantin L. Metlov}
\affiliation{Donetsk Institute of Physics and Technology, 72 R. Luxembourg str., Donetsk, Ukraine 83114}
\email{metlov@fti.dn.ua}
\date{\today}

\begin{abstract}
Here we consider an interplay between the topology of the magnetization texture (which is a topological soliton, or Skyrmion) in a planar magnetic nano-element and the topology of the element itself (its connectivity). We establish the existence of a set of constraints, coupling these topologies, which are specific for multiply connected elements and are absent in simply-connected case. As an example, a specific constraint is derived for a case of planar ring magnet, relating the angular positions of magnetic vortices and antivortices inside. We analyze the recent experimental data on the vortex magnetic domain walls in the ring to validate our findings.
\end{abstract}
\maketitle

Planar magnetic nanoelements are now quickly adopted for various applications of magnetism. Most notably MRAM\cite{S09}, STT-MRAM\cite{KW15} and patterned magnetic recording media\cite{R01}, but also as an integral part of other spintronic devices, such as nano-scale microwave generators\cite{DGGCKFKKYY10}, magnetic tunnel transistors\cite{J03} and other devices built on top of spin valves and magnetic tunnel junctions. This is mainly thanks to their compatibility with the existing nano-scale patterning technologies, such as ultra-violet and E-beam lithography. They are also extensively studied using numerical methods of micromagnetics.

However, the experiments (including numerical experiments) provide only with the knowledge of how a particular system behaves under a particular conditions, it is the task of theory to generalize the experiments to make the general statements about the nature and properties of particular phenomena. While such general statements in physics always involve approximations (and, therefore, always have limited applicability), they are tremendously useful for planning the experiments, devices, and for a general outlook what can be expected in the studied systems and what, on the contrary, is unexpected and, therefore, deserves special attention.

In this work we construct such general statements about properties of magnetization configurations in multiply-connected planar magnetic nanoelements. While the magnetization textures of simply connected elements are relatively well studied, the analytical theory of such a textures in multiply-connected elements have appeared only recently\cite{BM15}. It expresses the magnetization textures with the help of analytical functions of complex variable and, in particular, the Schottky-Klein prime functions\cite{Schottky1887,Klein1890}. These functions are not easy to evaluate\cite{CrowdyMarshall2007}, which somewhat diminishes the value of having such a compact analytical expressions. Yet, such a representation opens a possibility for symbolic analysis, making it possible to express some of the properties of magnetization textures in the form of simple constraints. These constraints, which have a topological nature and are directly related to the connectivity of the planar magnetic nano element, are the main subject of this work. While the analysis, leading to them, is approximate, and the observed (measured or computed) magnetization textures may differ locally from what is predicted by our simple analytical theory, it can be hoped that an intricate interplay of the topology of the magnetization textures (which are magnetic topological solitons, or Skyrmions) and the topology of the magnetic nano-element is captured by the present analysis well.

In the following we will briefly remind the reader of the approximate mapping of the problem of the magnetization textures of planar magnetic nanoelements to the complex analysis\cite{M10}, which is our starting point. Then we will derive the topological constraints for the case of doubly-connected planar nano-element (a ring). Finally, we analyze recent experimental data on the dynamics of head-to-head magnetic vortex walls in a ring\cite{Bisig2013VW} and check the established constraints. We discuss the applicability of our results and draw the conclusions at the end.

The approach is based on representation of a wide class of trial functions for the magnetization distributions in planar soft ferromagnetic elements at remanence by functions of complex variable\cite{M10}. It employs the magnetic energy terms hierarchy with the ferromagnetic exchange as the strongest and the energy of the magnetostatic volume charges as the weakest contributions. This hierarchy is well defined for the elements of sub-micron size, which are the main subject of this work. To build the trial functions, the energy terms are minimized sequentially from strongest to weakest, with each step building on top of the previous one and selecting the functions, which, additionally, minimize the next weaker energy term\cite{M10}. The exact solution in this picture corresponds to {\em simultaneous} minimization of all energy terms and is not available at the moment for the considered non-ellipsoidal nanomagnets. Nevertheless, the approximate expressions for the magnetization distributions, following from the described sequential minimization framework, are very convenient for obtaining useful closed form analytical results with excellent agreement to experiments \cite{MG02_JEMS,GHKDB06,M06,ML08,M13.dynamics}. 

Specifically, the Cartesian components of the resulting magnetization distributions are expressed as a stereographic projection
\begin{eqnarray}
 \label{eq:mxmymz}
 \umx+\imath \umy & = & \frac{2 w}{1+w\overline{w}} \\
 \umz & = & \frac{1-w\overline{w}}{1+w\overline{w}},
\end{eqnarray}
where the overline denotes complex conjugation and the coordinate system is chosen in such a way that $X$ and $Y$ coordinates define a point on the element's face (having an arbitrary shape ${\cal D}$). The elements are assumed to be thin enough so that $Z$-dependence of the magnetization (across the element's thickness) can be ignored. The representation ensures that the length of the reduced magnetization vector $|\vec{m}|=|\vec{M}|/\uMs=1$ is a constant at every point inside the magnet, where $\uMs$ is the ferromagnet's saturation magnetization. The function $w = w(z, \zc)$ is a complex function of the complex variable $z=X + \imath Y$, $z \in {\cal D}$, which is not necessarily analytic, but can be expressed via another analytic function $f=f(z)$ as
\begin{equation}
  \label{eq:sol_SM}
  w(z,\overline{z})=\left\{
    \begin{array}{ll}
      f(z)/e_1 & |f(z)| \leq e_1 \\
      f(z)/\sqrt{f(z) \fc(\zc)} & e_1<|f(z)| \leq e_2\\
      f(z)/e_2 & |f(z)| > e_2
    \end{array}
    \right. ,
\end{equation}
where $e_1$ and $e_2$ are two real scalars, satisfying $0<e_1<e_2<\infty$. The function $f(z)$, in turn, is the solution of the Riemann-Hilbert boundary value problem of finding the analytic function with no normal components on the boundary of the domain ${\cal D}$. The solution of this boundary value problem usually depends on a number of real and complex constants, which, together with $e_1$ and $e_2$, serve as Ritz parameters for final magnetic energy minimization, which includes all the energy terms irrespectively of their place in the hierarchy.

For example, in the simplest case of a single vortex in a circular disk, the solution of the above-specified Riemann-Hilbert problem is $f(z)=\imath z c + A - \overline{A} z^2$, with $c$ and $A$ being real and complex scalars respectively\cite{M01_solitons2}. It is possible to generalize this solution for the case of multi-vortex states\cite{M10} in which case it depends on additional scalar parameters, related to the positions of vortices and antivortices on the element's face.

For multiply-connected regions the multi-vortex (and antivortex) solution $f(z)$ corresponds to a \emph{real} meromorphic differential $d\omega=\ud z/f(z)$. This representation highlights the conformal invariance of the problem, which allows to relate the magnetization distributions in one domain $\cal D$ to those in the other, say, a canonical one. Thus, for considering all the doubly-connected domains, which will be our prime example in this paper, it is sufficient to study only a concentric ring (or annulus), which is the canonical domain for the doubly-connected case into which all other doubly-connected domains can be conformally mapped. 

The annulus, which we denote as $A:=\{z\in\mathbb{C}:\quad \rho\le|z|\le 1\}$, is defined by its inner radius $0<\rho<1$. If our original doubly-connected domain ${\cal D}$ is not an annulus, the parameter $\rho$ is called the conformal modulus and is uniquely defined in the process of conformal mapping, otherwise it can be arbitrary. Suppose we have a certain meromorphic function $f(z)$, which is a solution of the Riemann-Hilbert problem (with no normal components to the boundary of the annulus) and has a number of topological singularities: isolated zeros and poles, which may also be located at the boundary.

\begin{figure}[tb]
\centering
\includegraphics[width=\columnwidth]{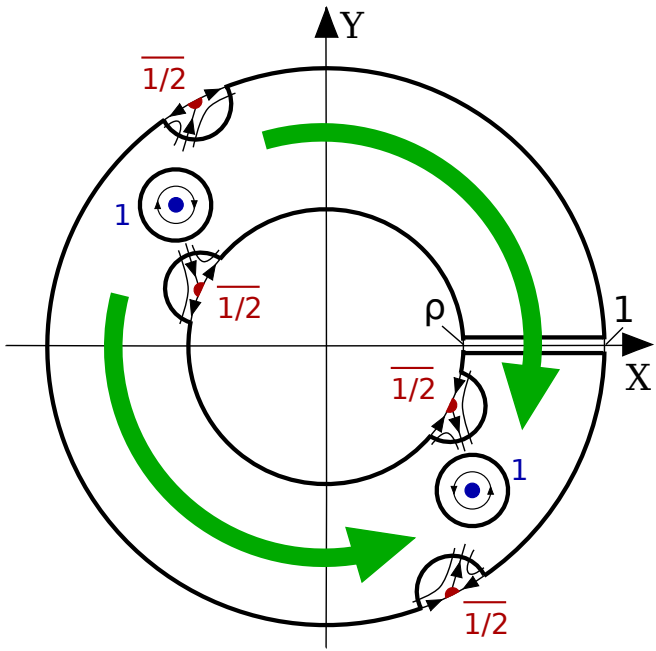}
\caption{\label{fig:contour}Integration contour in the equation (\ref{eq:zero}) for the case of two head-to-head 180\textdegree{} vortex walls in a ring. Two bold arrows show the general direction of the magnetization inside two magnetic domains. The topological singularities (vortices and antivortices) are shown schematically with their centers marked by bold dots (and half-dots if they are on the boundary). The vortices are labeled with $1$ and the half-antivortices at the boundary with $\overline{1/2}$.}
\end{figure}

Let us now find a specific form of such a constraint for the annulus $A$. We know that there exists a real holomorphic differential $\ud \eta = \imath \ud z/z$ without zeros and poles in $A$. Once we divide the differential $\ud \eta$ by $\ud \omega$ we get a meromorphic function $v(z)=\imath f(z)/z$, which is real on the boundary circles of the annulus and has zeros and poles exactly at the positions of vortices and antivortices of the original magnetization distribution. Slightly rotating the annulus, we can assume that $v(z)$ has no zeros and poles at the straight line segment $[\rho, 1]$. We then have the following identity:
\begin{eqnarray}
 0 & = & \iint\limits_{A-(v)} \ud\,\log{}(v(z))\wedge \ud{}\,\log{}(z) = \nonumber \\
& & -\oint\limits_{\partial(A-(v)-[\rho,1])}\log{}(z)~ \ud{}\,\log{}(v(z)), \label{eq:zero}
\end{eqnarray}
where $\wedge$ denotes the wedge product. The integration in the first integral goes over the annulus $A$ with excluded small disks around zeros and poles of the function $v(z)$ denoted by $(v)$, it is equal to zero because $\ud z \wedge \ud z = 0$. The integration contour for the second integral  is illustrated in Fig.~\ref{fig:contour}.
The second equality is due to the Poincar\'e-Stokes formula (integration by parts). We have to  cut the
annulus (along the segment $[\rho, 1]$) to single out the branch of $\log{}(z)$. 

On the boundaries of the small discs, including those centered at the boundary of the annulus, the integral is calculated by the residues formula and yields
\begin{equation}
 2\pi \imath \sum\limits_{z\in (v)} m(z)\log(z),
\end{equation}
where $z$ is the coordinate of each zero and pole, $m(z)$ is the multiplicity of the zero/pole $z$ of the function $f(z)$, negative once $z$ is a pole and taken with the weight $1/2$ if $z$ is on the boundary.

The integral over the boundary circles with excised symmetric vicinities of the boundary
vortexes is purely imaginary. And, finally, the integral over the banks of the slit $[\rho,1]$ corresponds to integrating the additive jump of the integrand across the banks, which is $- 2 \pi \imath \ud\, \log(v(z))$, along the slit. Since $v(z)$ is real at the slit endpoints, the real part of this integral lies in the lattice $2 \pi^2 \mathbb{Z}$, where $\mathbb{Z}$ is the set of integers. Thus, considering only the real part of the identity (\ref{eq:zero}) and dividing both sides by $2\pi$ we get the constraint
\begin{equation}
\Arg(f)_+ -\Arg (f)_- \in\pi\mathbb{Z}, \label{eq:constraint}
\end{equation}
where $\Arg(f)_\pm$ is the sum of the arguments of all zeros/poles of $f(z)$ in the annulus, counted as fractions of $2\pi$, taken with their multiplicities and weighted by $1/2$ if the pole/zero lies at the boundary of the annulus. In other words: {\em the sum of the azimuths of the vortices (counted from the center of the annulus as fractions of $2\pi$) is equal to the sum of azimuths of the antivortices  modulo $\pi$}.

Similar, but more sophisticated, constraints exist also for the regions of the higher connectivity, which we plan to explore in our future work. They essentially stem from Abel's theorem for algebraic functions, which implies there should be $c-1$ similar real constraints for positions of zeros and poles, where $c$ is the connectivity of the region. In the simply-connected regions vortices and antivortices do not have such restrictions on their positions.

Another topological constraint we have already explored in our previous work\cite{BM15} connects the {\em numbers} of vortices $N_\mathrm{V}$  and antivortices $N_\mathrm{A}$ (also counted with their multiplicities and taken with the factor $1/2$ if the vortex/antivortex lies at the boundary) in a multiply-connected region by the relation $N_\mathrm{A}-N_\mathrm{V} = c - 2$, where $c$ is again the connectivity of the region. It is not specific for a particular magnetization texture configuration and follows from Hopf-Poincar\'e theorem about the sum of indexes of singular points of a vector field.

Since all our arguments here and in Ref.~\onlinecite{BM15} are reversible, the inverse statement is also true: in a specific multiply-connected region for given set of singularities, satisfying both topological constraints, there always exists a unique (up to a dilation) solution $f(z)$ of the Riemann-Hilbert problem (magnetization texture).

To verify this result one may check that it holds exactly against all the magnetization distributions in a ring, shown as illustrations in Ref.~\onlinecite{BM15}. Note that, while the resolution of pictures there is rather high, we have performed verification of the constraint (\ref{eq:constraint}) using numerical root finding technique. The constraint is satisfied exactly up to the machine precision (around 10 decimal digits). The experimental verification is more difficult due to the approximate nature of the theory and also due to the difficulty determining the exact topological singularity positions in the relevant nano-scale systems. 

\begin{figure}[tb]
\centering
\includegraphics[width=\columnwidth]{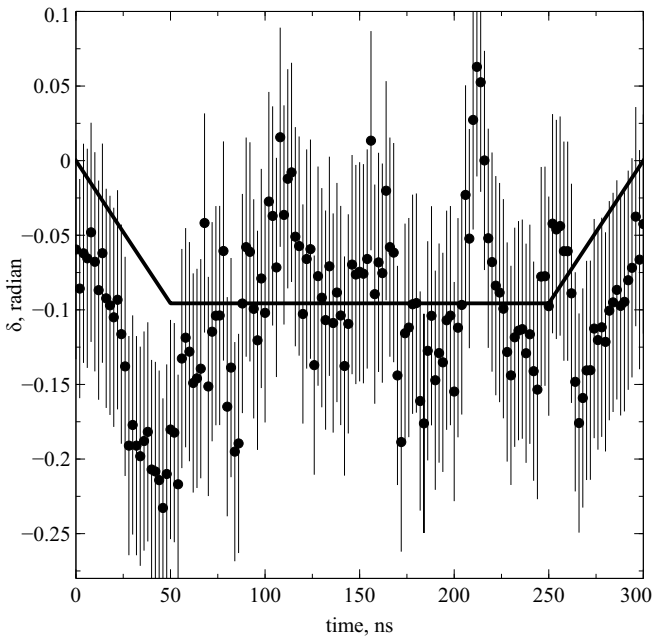}
\caption{\label{fig:angles}The sum of azimuth of vortices and antivortices $\delta$ as a function of time during the complex dynamical evolution of two vortex walls in a ring, extracted from the experimental data of Ref.~\onlinecite{Bisig2013VW}. The solid line shows the magnitude of the external field during the evolution in arbitrary units taken with the minus sign for easier comparison. The field is increasing, performing a  360\textdegree{} rotation in the ring plane, then decreasing.}
\end{figure}

Thankfully, there is a recent experiment\cite{Bisig2013VW} on vortex domain wall dynamics in a microscale permalloy ring, which allows to resolve the positions of vortices (done in the cited paper) and antivortices (done by us) as functions of time, corresponding to the time-varying amplitude and direction of the external field. The computed phase difference $\delta = (\Arg(f)_+ -\Arg (f)_-) \bmod \pi$ is shown in Fig.~\ref{fig:angles}.
The experimental images were acquired stroboscopically, starting from a remanent state, then increasing the magnetic field amplitude, then performing the full rotation of the field (of constant magnitude) in the ring plane, then decreasing the field back to zero, after which the cycle was repeated. They are published as a movie in a supplemental materials to Ref.~\onlinecite{Bisig2013VW}. The frames of the movie were processed in the original publication by placing the rather large bold dots (with the angular size of about 0.1 radian) on top of magnetic vortices. We have placed a similar sized dots on top of magnetic antivortices and recorded their positions to compute the phase difference $\delta$ and plot it in the Figure. The updated movie with all the topological singularities marked is available as a supplemental material\cite{suppmoviefileAPS}. The scatter of the points is substantial and has both random (due to the pinning) and systematic (due to the size of the dots, placed on top of the image) contributions. The random error was estimated by assuming that at fixed magnitude of the external field the scatter is entirely due to the random factors, thus, the random error of $\sigma_R=0.0544$ radian was computed as a standard deviation of the $\delta$ over the time range from $50$ to $250$ nanoseconds. The systematic error $\sigma_S=0.0493$ radian was taken to be a half of the angular size of the vortex-position dot. The error bars of $\sigma=\sqrt{\sigma_R^2+\sigma_S^2}\simeq0.08$ radian were estimated by combining these two errors.

The original constraint was derived for the state of remanence (zero applied field). This corresponds to two values of $\delta=-0.06\pm0.08$ and $\delta=0.04\pm0.08$ at $0$ and $300$ nanoseconds. From this we may directly conclude that the experiment supports our result. However, there is much more information in the data which merits the discussion.

First of all, we can see that $\delta$ is mostly negative, which means that the antivortices generally move ahead of vortices during the field cycle. This can be explained by the fact that they are closer to the ends of the magnetic domains, which are directly rotated by the field. The vortex is in between of antivortices and has less traction. The smaller traction and the intrinsic pinning on the material and ring thickness imperfections makes the vortex lag behind, which is just a manifestation of the hysteresis. The deeper is the pinning potential, the larger is the lag. Thus, the presence of pinning implies that $\delta$ may assume not just a single value of $0$ (as in a perfect material), but a certain range of values, which are selected based on the magnetic history of the sample. For cyclic magnetization process of Ref.~\onlinecite{Bisig2013VW} this results in a constant lag.

Another feature of the experimentally measured phase difference in Fig.~\ref{fig:angles} is that it somehow follows the magnitude of the external field: increasing when field increases, staying mostly constant (with large fluctuations) when the field magnitude is a constant, and decreasing when the field is decreased. Let us remember that the constraint with $\delta=0$ (e.g. with no lag of vortices) was derived for a perfect material in zero applied field, assuming the energy hierarchy with the exchange on top and the energy of side and volume magnetic charges at the bottom (of which the energy of side charges was completely eliminated). Once the in-plane magnetic field is taken into account, the Zeeman energy must also find its place in this hierarchy. We know that the field forces the ring to acquire the side magnetic charges (increasing magnetization component, normal to the ring boundaries). Supposing the field is low enough that the exchange interaction is still dominating, we can retain all the analyticity properties of the function $f(z)$, but have to reject the assumption that $v(z)$ is real at the boundary of the ring. This means that the contour integral along the boundary circles with excised symmetric vicinities of the boundary vortexes is not purely imaginary anymore, but has a certain real part, which we will denote as $\delta_h$. In the case of symmetric circular ring this real part may depend only on the magnitude of the external field $h$. Thus, the presence of the external field modifies the constraint by adding the $\delta_h$ term. The computation of this term as well as of the effects of pinning may require substantial extension of the model and is beyond the scope of the present paper. Nevertheless, the presence of both the lag (the constant at $h=0$ contribution to $\delta$) and the effect of the external field (the field-magnitude-dependent contribution in case of circular ring) does not dismiss the observation that the azimuth of vortices and antivortices during a very complex dynamical evolution of vortex walls in a ring \cite{Bisig2013VW} are bound by a constraint due to the fact that the ring is a doubly-connected region.

Thus, we have shown that in the multiply-connected planar sub-micron and micron ferromagnetic elements at remanence, there can be topological constraints on vortex and antivortex positions, which arise entirely due to high connectivity of the region. We have derived the specific form of the constraint for the canonical doubly-connected domain, which is a concentric circular ring. The constraint is verified and holds exactly against our earlier analytical solutions\cite{BM15}. Its existence is also confirmed by the recent experiment on the dynamics of two vortex walls in the ring\cite{Bisig2013VW}, which additionally demonstrate that restrictions on the vortex movement exist even beyond our original assumption of zero applied field. There are no analogs of such constraints in simply-connected elements, where the positions of vortices and antivortices are determined by the least important terms in the energy hierarchy.

Higher element connectivity opens another, largely unexplored, avenue for design of spintronic devices, since it allows additional means of controlling the magnetization of functional magnetic elements and its dynamics. We hope that understanding the topological restrictions of the vortex/antivortex placement and movement will help in this very interesting and promising endeavor.

\begin{acknowledgments}
The support of the Russian Foundation of Basic Research under the project {\tt RFBR 16-01-00568} is acknowledged. 
\end{acknowledgments}

%

\end{document}